\documentclass[%
 aip,,
 amsmath,amssymb,
 reprint,%
]{revtex4-2}

\usepackage{amsmath}
\usepackage{graphicx}%
\usepackage{dcolumn}%

\usepackage[utf8]{inputenc}
\usepackage[T1]{fontenc}
\usepackage{lmodern}  %
\usepackage{mathptmx}
\usepackage{etoolbox}
\usepackage{booktabs}

\usepackage[bookmarksnumbered=true]{hyperref} 
\hypersetup{
     colorlinks = true,
     linkcolor = blue,
     anchorcolor = blue,
     citecolor = blue,
     filecolor = blue,
     urlcolor = blue}

\usepackage{xcolor}
\usepackage{siunitx} %
\usepackage{nicefrac}

\newcommand{\alert}[1][]{}

\newcommand{\equref}[2][]{Eq.\,\ref{#2}#1}

\newcommand{\figref}[2][]{Fig.\,\ref{#2}#1}
\newcommand{\Figref}[2][]{Figure\,\ref{#2}#1}
\newcommand{\tabref}[2][]{Tab.\,\ref{#2}#1}

\usepackage{lineno}

\makeatletter
\let\@fnsymbol\@fnsymbol@latex
\@booleanfalse\altaffilletter@sw
\makeatother

\begin{document}

\title[ZGV nano-imaging]{Frequency domain laser ultrasound microscopy for nanometric layer thickness imaging with GHz elastic plate resonances}

\author{Martin Ryzy}
\affiliation{Research Center for Non-Destructive Testing GmbH, Linz, Austria}

\author{Guqi Yan}
\email[Corresponding author: ]{guqi.yan@recendt.at}
\affiliation{Research Center for Non-Destructive Testing GmbH, Linz, Austria}

\author{Istv\'{a}n Veres}
\affiliation{Qorvo, Inc., Apopka, FL 32703 USA}

\author{Thomas Berer}
\affiliation{Qorvo, Inc., Apopka, FL 32703 USA}

\author{Ivan Ali\'{c}}
\affiliation{Research Center for Non-Destructive Testing GmbH, Linz, Austria}

\author{Clemens Grünsteidl}
\affiliation{Research Center for Non-Destructive Testing GmbH, Linz, Austria}

\author{Georg Watzl}
\affiliation{Research Center for Non-Destructive Testing GmbH, Linz, Austria}

\author{Georg Gramse}
\affiliation{Johannes Kepler University, Biophysics Institute, Linz, Austria}

\author{Susanne Kreuzer}
\affiliation{Qorvo, Inc., Apopka, FL 32703 USA}

\begin{abstract} %
	Nanometric layer thickness imaging is crucial for fundamental research and characterization of micro fabricated devices. Here, we assess the potential of a non-contact opto-acoustic frequency domain laser ultrasound (FreDomLUS) microscopy technique for imaging nanometric thickness variations via GHz zero-group velocity (ZGV) elastic plate resonances. The method exploits the ZGV's lateral energy confinement that leads to sharp resonance peaks which can be effectively probed with the FreDomLUS technique at GHz acoustic frequencies. For demonstration purposes we introduced sub-\SI{10}{nm} height variation patterns in the topmost layer of solidly mounted bulk-acoustic wave resonators with a design frequency of around \SI{1.7}{GHz}. They are raster-scanned to retrieve ZGV-frequencies from local acoustic spectra as a contrast quantity for imaging. We show how to retrieve quantitative height information by numerically calibrating the factor which inversely relates ZGV frequency change with the layer thickness change. Height variations in stacks with nominal thickness changes of \SI{8}{nm}, \SI{4}{nm}, and \SI{1}{nm} can be resolved and indicate sub-nanometer depth resolution capabilities. The lateral resolution is studied by measuring the method's step edge function and it is found to be in the micrometer range. Atomic force microscopy imaging is used to validate the results.
\end{abstract}

\maketitle

\section*{Introduction}
Nanometric surface topography and layer thickness imaging is of great importance for basic research, device characterization, and quality control. Nowadays a multitude of well-established methods with nanometric depth resolution like different flavors of scanning probe microscopy, scanning electron microscopy and optical profilometry techniques are available.\cite{d_j_whitehouse_surface_1997, el_rifai_imaging_2003} For the characterization of optically opaque materials, ultrasonic waves have proven useful due to their high penetration depth. For example, scanning acoustic microscopes (SAMs) are widely used for surface and sub-surface layer characterization of integrated circuits and microfabricated devices. The technique is based on evaluating reflected longitudinal acoustic waves which are launched and recorded by a piezoelectric contact-transducer. To achieve high lateral and axial resolution, GHz frequency acoustic waves are usually used. However, they are often strongly absorbed in solids which limits the in-depth resolution to around \SI{1}{\micro\metre}.\cite{lemons_acoustic_1974,maev_acoustic_2008,bertocci_scanning_2019} For thin film characterization ultrafast opto-acoustic techniques can be used which also utilize acoustic bulk waves, but with even higher frequencies beyond \SI{10}{GHz}. Here, strong acoustic loss limits the application of the technique usually to film thicknesses in the nanometer range.\cite{PhysRevLett.110.095503,Grossmann_2017}

In plate-like structures, laterally guided elastic plate-waves (termed Lamb-waves) emerge in addition to bulk- and surface acoustic waves, if the bulk wavelength is in the order of the plate thickness or larger.\cite{Rayleigh1889,Lamb1889,graff1991_p431} Due to their strongly dispersive nature, counter-intuitive resonance phenomena can be observed: at certain frequencies, non-propagating long-ringing resonances with a finite wavelength appear. They are characterized by a vanishing group velocity and thus called zero group velocity (ZGV) resonances.\cite{tolstoy_wave_1957,Gibson2005,Prada2005} When they are excited with a local acoustic source with a lateral extent smaller or comparable to the ZGV-wavelength, their acoustic energy will be trapped underneath the source. As a result, these resonances dominate the local response spectrum, where they appear as sharp peaks in acoustic spectra and they can be considered as local in plate direction.\cite{Prada2005,Bernard2001,Prada2008a} While the concentration of acoustic energy around a certain frequency facilitates ultrasonic measurements at high frequencies,\cite{Grunsteidl2020} where materials are typically highly absorptive, the lateral spatial confinement enables local material characterization: for instance, measuring ZGV-resonance frequencies can be used to gain local information about elastic properties\cite{Grunsteidl2018} and to image thickness variations of bare\cite{Grunsteidl2018,Clorennec2010,Balogun2011} or coated plates,\cite{holland_air-coupled_2003,holland_high_2004,Ces2011} as they are governed by the electro-elastic properties and thicknesses of the plate or its individual layers. This can be done by all optical contactless laser ultrasonic (LUS) techniques: here, a focused excitation laser (pulsed or intensity modulated) provides a local ultrasound source by thermoelastic energy conversion, and a second laser probes the samples' surface displacement or velocity by interferometric means for instance.\cite{hutchins_ultrasonic_1988,Scruby1990_v_p,davies_laser-generated_1993,spytek_non-contact_2023,murray_high-sensitivity_2004,kokkonen_laser_2009} With this technique, a multitude of ZGV-resonance based methods for material characterization have been developed in the last two decades, ranging from aforementioned thickness measurements\cite{Grunsteidl2018,Clorennec2010,Balogun2011,Ces2011} and determination of elastic properties,\cite{Grunsteidl2018,ces_characterization_2012,baggens_poissons_2015,Grunsteidl2016,yao_characterization_2019,Watzl2022} to characterization of adhesion,\cite{Mezil2015,Hode2020,Spytek2020} metal fatigue\cite{Yan2018,Yan2020} nano-porosity,\cite{ahn_elastic_2014,Thelen2021} and acoustic loss.\cite{Prada2008,Laurent2014,Yan2021,Ryzy2023} First applications were restricted to MHz frequencies,\cite{Prada2005,Clorennec2010,Clorennec2007} but recently the range was extended to the GHz regime.\cite{Ryzy2023,berer_determination_2019,Xie2019,grunsteidl_evaluation_2020}

Bulk acoustic wave (BAW) resonators are micro-electro-mechanical systems (MEMS) which are used in radiofrequency (RF) filters of mobile communication devices. They typically consist of multiple sub-micron layers of different materials, with a piezoelectric layer for electro-elastic energy conversion.\cite{Lakin1981,Lakin2003,Miller2015} Their electro-elastic response can be characterized by impedance spectroscopy or interferometric measurements for instance.\cite{kokkonen_laser_2009,kokkonen_direct_2010,kokkonen_measurement_2011} In both cases the devices are electrically excited with an RF signal applied to their planar electrodes. Thus, their entire surface area vibrates, and the gathered information must be considered as non-local. However, there is a strong desire to get local information about the electro-elastic properties of BAW resonators, to study edge effects like lateral energy leakage and to optimize microfabrication processes. As ZGV plate-resonances have been shown to also exist in multi-layer plates\cite{kuznetsov_guided_2023} and electro acoustic resonators,\cite{Yantchev2011,Caliendo2017,caliendo_design_2018} a local, ZGV-resonance based characterization method is a promising approach for this purpose. 

In this paper, we demonstrate ZGV-resonance based sub \SI{10}{nm} topographic layer thickness imaging with a non-contact frequency domain laser ultrasound (FreDomLUS) microscopy system. For this purpose, we produced patterns of varying thicknesses in the topmost layer of solidly mounted \SI{1.7}{GHz} BAW resonators and raster scanned their acoustic response around their ZGV center-frequency. From that, images were generated and compared to atomic force microscope (AFM) images. The lateral resolution of the FreDomLUS technique was quantified by measuring its edge spread function.

\begin{figure*}[b!htp]
	\centering
	\includegraphics[width=\textwidth]{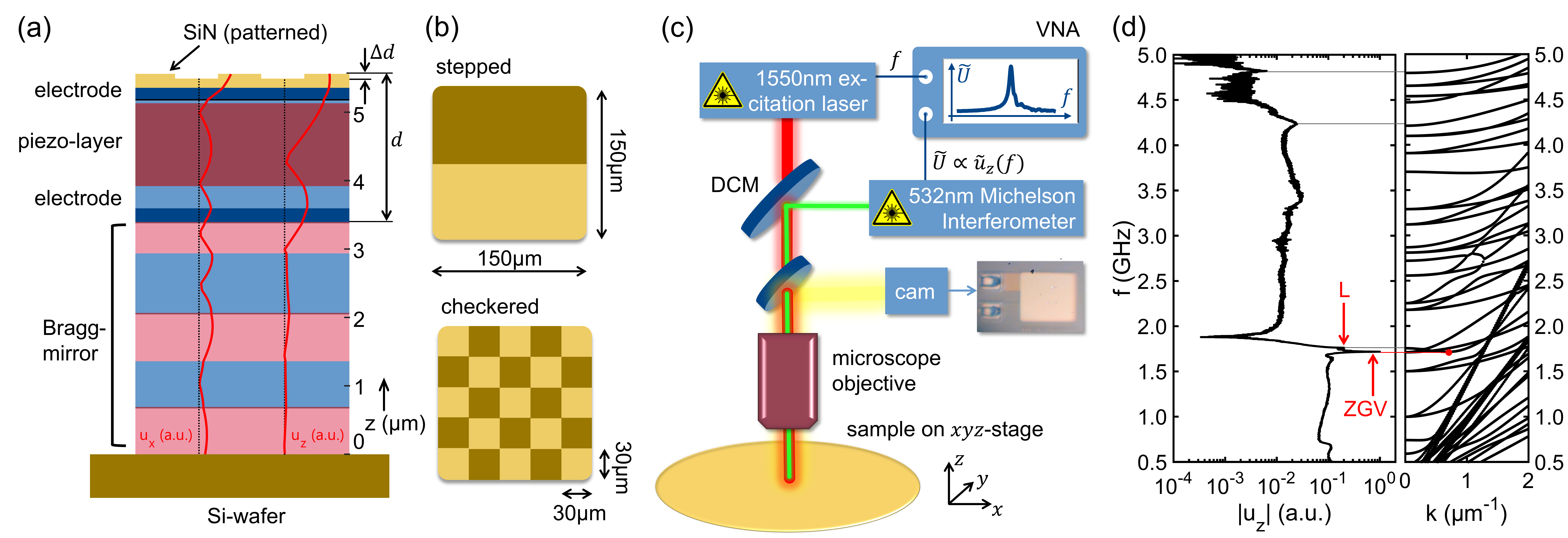}
	\caption{\textbf{Samples and Setup.} (a) Sketch of the cross section of the studied BAW-resonator samples. They consist of a piezoelectric Sc$_x$Al$_{1-x}$N-layer ($x\!=\!0.09$) sandwiched by two metallic electrodes placed on an acoustic Bragg mirror. The topmost \SI{205}{nm} thick SiN-passivation-layer is structured with pits of depth $\Delta{}d$ that are arranged in either a stepped (b, top) or checkered (b, bottom) geometry. Note that the lateral dimensions of the structures (\SI{150}{\micro\metre}) are large compared to the total thickness of the layer-system (\SI{5.562}{\micro\metre}). The red-solid lines are calculated mode-shapes at the ZGV-point (c) Sketch of the FreDomLUS-microscopy setup used for the measurement of acoustic response spectra. (d) Broadband GHz-acoustic response spectrum (left) from an unstructured region of a BAW-device sample and calculated plate-wave dispersion curves (right). The strongest peak in the spectrum is identified as zero-group velocity resonance (labeled as ZGV).\label{fig:Figure_ArticleVersion_1}}%
\end{figure*}

\section*{Results}
\subsection*{Laser-based measurement of ZGV-resonances in BAW-resonator samples}
We studied solidly mounted bulk-acoustic wave resonator stacks as depicted in \figref[(a)]{fig:Figure_ArticleVersion_1}. They are composed of multiple-layers deposited on a Si-substrate: A \SI{2.73}{\micro\metre} thick acoustic Bragg-mirror consisting of five alternating layers of SiO$_2$ and W, followed by a \SI{1.2}{\micro\metre} Sc$_x$Al$_{1-x}$N-layer (with $x\!=\!0.09$; denoted as ScAlN9) piezoelectric layer between two sub-micron metallic-electrodes. With this piezo-thickness they have an acoustic breathing resonance at around \SI{1.7}{GHz}. The stacks are quadratic with a lateral dimension of \SI{150}{\micro\metre} and are passivated with a \SI{205}{\nano\metre} thick SiN-layer. In these top layers height-change patterns were introduced, either in the form of a single step ('stepped resonator') or as a checkerboard pattern ('checkered resonator') with a period of \SI{30}{\micro\metre} (\figref[(b)]{fig:Figure_ArticleVersion_1}). Several wafers were produced, each having a different nominal depth of the ablated regions of either \SI{8}{nm}, \SI{4}{nm}, or \SI{1}{nm}.

For recording acoustic response spectra we use a frequency domain laser ultrasound\cite{murray_high-sensitivity_2004, Ryzy2018a} system, which was previously optimized for GHz-frequency operation.\cite{Grunsteidl2020,Ryzy2023} It consists of an electro-absorption modulated diode laser (EML) with a wavelength of \SI{1550}{nm}, which is amplified to about \SI{250}{mW} by an erbium doped fiber amplifier (EDFA) and focused onto the sample with a microscope objective (see \figref[(c)]{fig:Figure_ArticleVersion_1}). The surface normal displacement is detected with a path-stabilized Michelson Interferometer (\SI{532}{nm} continuous wave Nd:YAG laser) connected to a vector network analyzer (VNA) for phase sensitive detection. A white light microscope provides a magnified top-view of the samples for aiming purposes. A mechanical $xyz$-translation stage is used for moving the samples relative to the co-aligned lasers and to spatially resolve acoustic response spectra for imaging.

The FreDomLUS-setup allows us to measure up to frequencies of about \SI{5}{GHz} with sufficient signal-to-noise ratio, as shown in the broadband acoustic response spectrum recorded on an unpatterned part of one of the resonators (\figref[(d)]{fig:Figure_ArticleVersion_1}). Several resonance peaks can be observed, which excellently match points of vanishing group velocity in theoretical dispersion curves $f(k)$ calculated with a spectral collocation method (SCM).\cite{kiefer_electroelastic_2025} The most pronounced one is the first ZGV-resonance at around \SI{1.7}{GHz}, followed by the first thickness stretch resonance a little higher in frequency (denoted as 'L' in \figref[(d)]{fig:Figure_ArticleVersion_1}).

\subsection*{ZGV-resonance frequency and thickness monitoring}
\begin{figure}[h!tbp]
	\centering
	\includegraphics[width=0.3\textwidth]{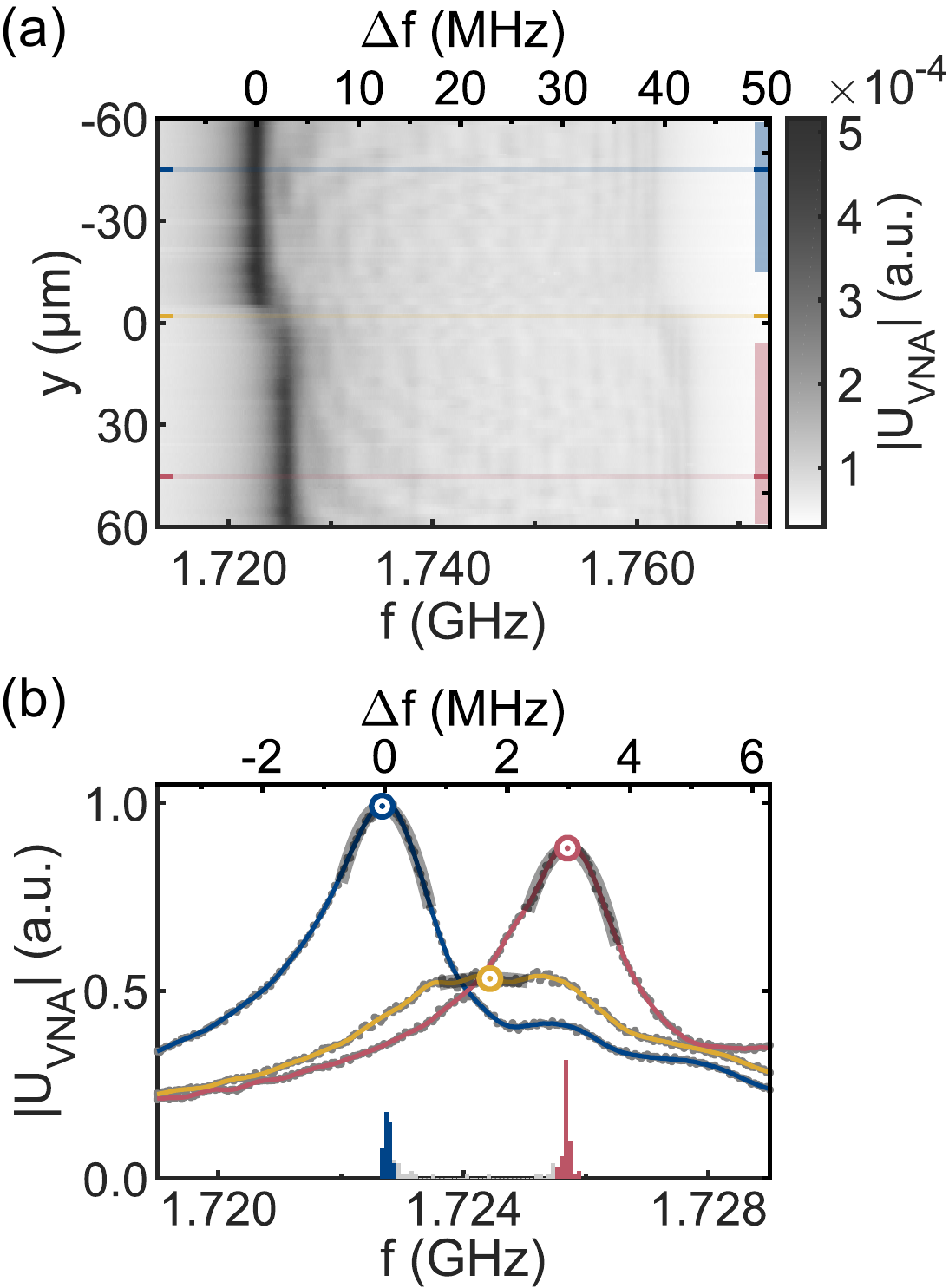}
	\caption{\textbf{FreDomLUS line scan on a 8\,nm stepped resonator.} (a) Narrowband acoustic spectra recorded along a line in $y$-direction. The dark region corresponds to the ZGV-resonance peak and clearly reflects the thickness step in the sample. (b) Example spectra taken from an unpatterned-region (blue), a patterned-region (with reduced thickness; red) and from the transition region (yellow). The locations of the spectra are indicated by horizontal lines in (a). The grey-dots are the original measured data-points, while the solid colored lines are smoothed by Savitzky-Golay filtering. The overlayed bars are histograms of extracted ZGV-frequencies taken from the unpatterned (blue) and patterened-region (red) which are marked by vertical thick lines in (a). \label{fig:Stepped_8nm}}%
\end{figure}

In (single-layer) plates, the resonance center frequencies are inversely proportional to the total plate thickness\cite{graff1991_p431}. Thus, even without knowing the plates elastic properties, the resonances can be used to monitor height variations via the relation\cite{Prada2005}
\begin{equation}
	\frac{\Delta f_\mathrm{ZGV}}{f_\mathrm{ZGV}} = -K \, \frac{\Delta d}{d},
	\label{eq:freqThick}
\end{equation}
where $K\!=\!1$. Here, $f_\mathrm{ZGV}$ is the ZGV-resonance center frequency of a plate with thickness $d$, and $\Delta{}f_\mathrm{ZGV}$ is the according frequency change if the plate thickness changes by $\Delta{}d$. For a plate composed of a substrate and a thin layer, the factor $K\neq1$, and it becomes dependent on the elastic properties and the densities of layer and substrate\cite{Ces2011}.
We expect a similar behavior for the multi-layer resonator stacks and thus recorded narrowband acoustic spectra in the range of the first ZGV-resonance in our samples. In the stepped samples, line scans in $y$-direction across the thickness step were performed as illustrated for the \SI{8}{nm} stepped-resonators in \figref[(a)]{fig:Stepped_8nm}. The figure clearly shows that, despite the tiny thickness change, the ZGV-peak (dark-region in the 2D-map) follows the step in the sample. Note that the signal-patterns at higher frequencies ($f\!>\!f_\mathrm{ZGV}$) can be attributed to standing waves which are formed by propagating modes near the ZGV-resonances (non-zero group velocity; see dispersion curves in \figref[(d)]{fig:Figure_ArticleVersion_1}) that are beeing reflected from the sample edges.\cite{Grunsteidl2020} \Figref[(b)]{fig:Stepped_8nm} shows three example spectra from different locations on the patterned surface: A spectra from an untreated region with original thickness (blue-line), one from an ablated region with reduced thickness (red-line), and a third from the transition region (the step; yellow-line). The locations are indicated by the horizontal colored lines in \Figref[(a)]{fig:Stepped_8nm}. The transition spectrum is clearly broadened and seems to be influenced by either resonance peaks (from the thick and thin region). To extract the ZGV-center frequencies (marked as large circles) for further processing, we  fitted the spectra by 2$^\mathrm{nd}$-order polynomials (thick semi-transparent grey lines).

\subsection*{Quantifying thickness variation and lateral resolution}
\begin{figure}[h!tbp]
	\centering
	\includegraphics[width=0.43\textwidth]{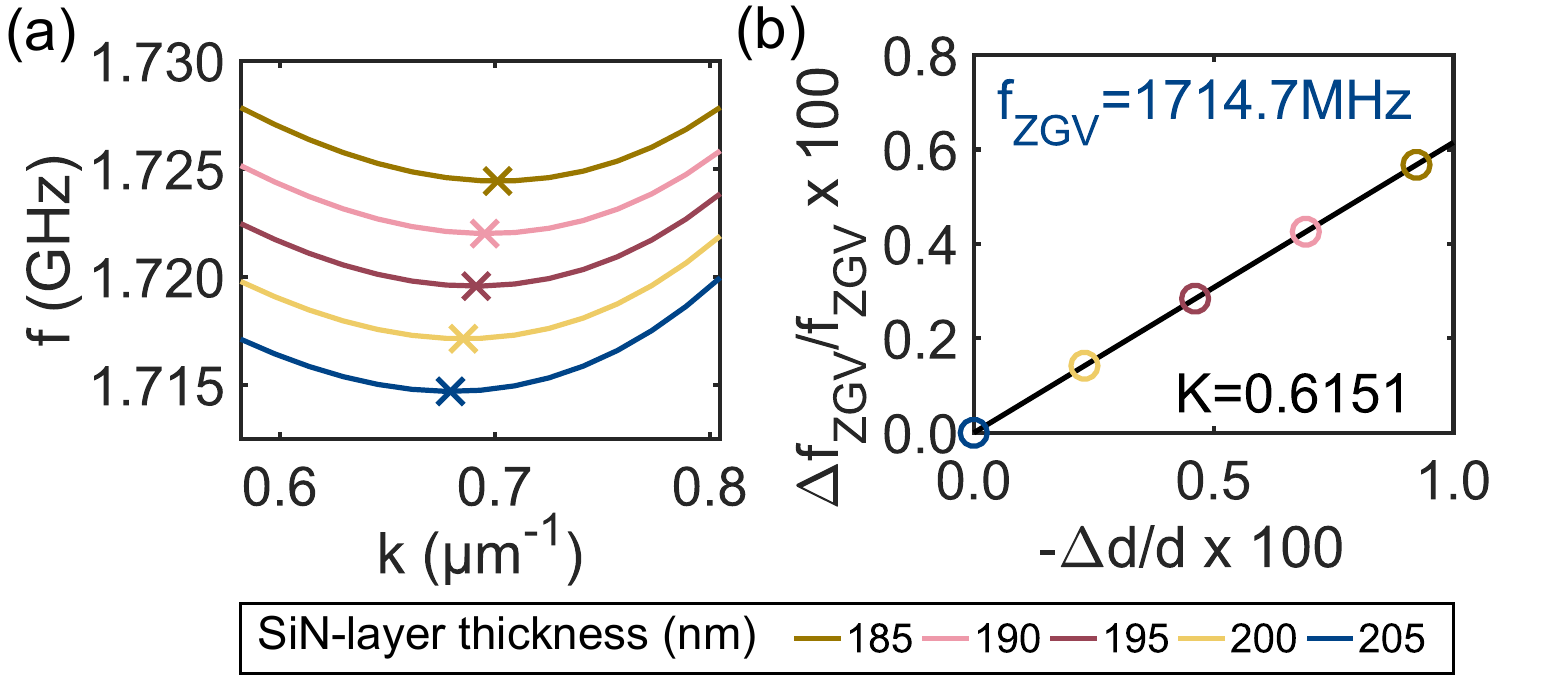}
	\caption{\textbf{Calibration of thickness model.} (a) Calculated BAW-stack dispersion curves in the region of the first ZGV-resonance with varying top SiN-layer thickness. The ZGV-points are marked by a cross. (b) Relative change of ZGV-frequency vs. relative thickness change. The reference thickness $d\!=\!\SI{2167}{nm}$ corresponds to the stack thickness without Bragg-mirror.\label{fig:calib_K_fit}}%
\end{figure}
\begin{figure*}[t!]
	\centering
	\includegraphics[width=1.0\textwidth]{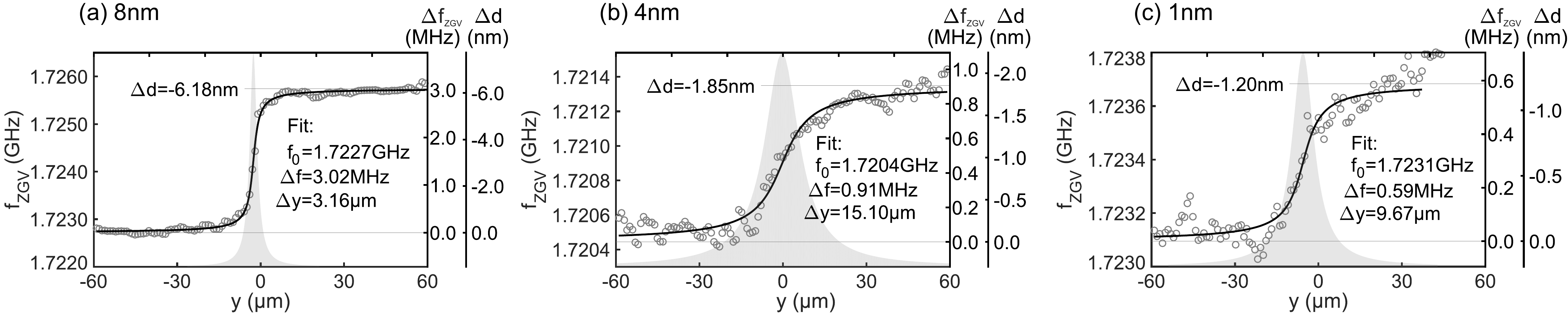}
	\caption{\textbf{Thickness variation from ZGV-measurements.} ZGV-resonance center-frequency variation (grey data points) from FreDomLUS-measurements in stepped resonators with different nominal step-heights of \SI{8}{nm} (a), \SI{4}{nm} (b), and \SI{1}{nm} (c). The additional y-axis on the right side of the plots illustrate the frequency change $\Delta{}f_\mathrm{ZGV}$ with respect to the unpatterned regions of the resonators and the estimated thickness variation $\Delta{}d$ according to \equref{eq:freqThick}. The solid black lines are fits of a step-edge function which has been used to quantify the total step-height $\Delta{d}$ and the lateral resolution $\Delta{}y$ of the method. The grey areas are the distribution functions related to the chosen step-edge function.\label{fig:steppedHeight}}%
\end{figure*}

We tested the validity of \equref{eq:freqThick} for the small thickness variations in the studied multi-layer stack by analyzing a series of dispersion curves calculated with a SCM-method.\cite{kiefer_electroelastic_2025} Here, the thickness of the topmost SiN-layer was varied in \SI{5}{nm} steps from \SI{205}{nm} to \SI{185}{nm}, and the ZGV-frequencies were extracted from the local minima of the ZGV-resonance branch in the dispersion curves (see \figref[(a)]{fig:calib_K_fit}). As illustrated in \figref[(b)]{fig:calib_K_fit}, a linear relation between relative frequency- and thickness-change (i.\,e, \equref{eq:freqThick}) also holds for this multi-layer system. By choosing a reference thickness of $d\!=\!\SI{2167}{nm}$, which corresponds to the stack without the Bragg-mirror (see \figref[(a)]{fig:Figure_ArticleVersion_1}), we obtain a proportionality factor of $K\!=\!0.615$. Note that we tested the validity of \equref{eq:freqThick} only for small thickness changes of the topmost layer ($\nicefrac{\Delta{}d}{d}\!\leq\!0.01$), and expect a deviation from the linear behavior for larger changes, as well as a different proportionality factor $K$ if a the thickness of a different layer in the stack varies.

\Figref{fig:steppedHeight} shows extracted ZGV-resonance frequencies from a \SI{8}{nm}, \SI{4}{nm}, and \SI{1}{nm} stepped BAW-resonator FreDomLUS line-scans. They clearly follow the step-profiles of the samples. The resonance frequency changes ($\Delta{}f_\mathrm{ZGV}$ shown on the right hand side y-axis of the plots) are around three orders of magnitude lower than the resoance center frequencies (MHz vs.\ GHz), and range from \SI{0.18}{\percent} (\SI{8}{nm}-case) to \SI{0.03}{\percent} (\SI{1}{nm}-case). To quantify the height-change, we generated an additional $\Delta{d}$-vertical-axes at the right hand sides of the plots in \figref{fig:steppedHeight} by applying \equref[]{eq:freqThick} with the previously calibrated $K$-value.

In order to measure the step-height and to estimate the lateral resolution of the method, we fitted a step-edge-function to the measured data. This is inspired by the knife-edge technique, which is often used to quantify the width of Gaussian laser-beams. We chose a scaled version of the cumulative distribution function of the Cauchy-distribution, which reads
\begin{equation}
	f_\mathrm{ZGV}(y) = f_0 + \Delta{}f \frac{1}{2} \left( 1 + \frac{2}{\pi} \arctan\left(\frac{2(y-y_0)}{\Delta{}y}\right) \right) \mathrm{.}
	\label{eq:CDF2}
\end{equation}
Here, the fit-parameter $f_0$ is the ZGV-frequency in the non-ablated region of the sample, $\Delta{}f$ the total frequency change caused by the step, $y_0$ the location of the step, and $\Delta{}y$ the full width at half maximum (FWHM) of the Cauchy distribution function. Both, the fits (solid black line) and the fit-parameters are shown in \figref{fig:steppedHeight}. From $\Delta{}f$ and \equref{eq:freqThick} the step-heights $\Delta{}d$ can calculated. They amount for \SI{6.18}{nm}, \SI{1.85}{nm} and \SI{1.2}{nm} for the different cases (see annotations in \figref{fig:steppedHeight}) and all differ from the nominal step heights of \SI{8}{nmn}, \SI{4}{nm}, and \SI{1}{nm}. If a perfectly abrupt step is assumed, the lateral resolution of the method may be quantified by the $\Delta{}y$ of the distribution function, which is illustrated as grey area in \figref{fig:steppedHeight}. It amounts to \SI{3.2}{\micro\metre}, \SI{15.1}{\micro\metre} and \SI{9.7}{\micro\metre} for the three cases, and is about three orders of magnitude larger than the measured step-heights. The compared to the thickness-sensitivity lower lateral resolution is related to the lateral extent of the used plate-resonances which is in the order of half the resonance wavelength.\cite{Prada2008a, ces_edge_2011} This should be approximately the same for all studied cases and amounts to $\nicefrac{\lambda}{2}=\nicefrac{\pi}{k}\approx\SI{4.5}{\micro\metre}$ ($k\!\approx\!\SI{0.7}{\per\micro\metre}$; see \figref[(a)]{fig:calib_K_fit}). However, only the $\Delta{}y\!=\!\SI{3.2}{\micro\metre}$ of the 8nm case is of comparable magnitude, while for the other two cases $\Delta{}y$ is larger. At this point it remains unclear why this is the case.

\subsection*{Quantitative height imaging}
\begin{figure*}[h!tbp]
	\centering
	\includegraphics[width=1.0\textwidth]{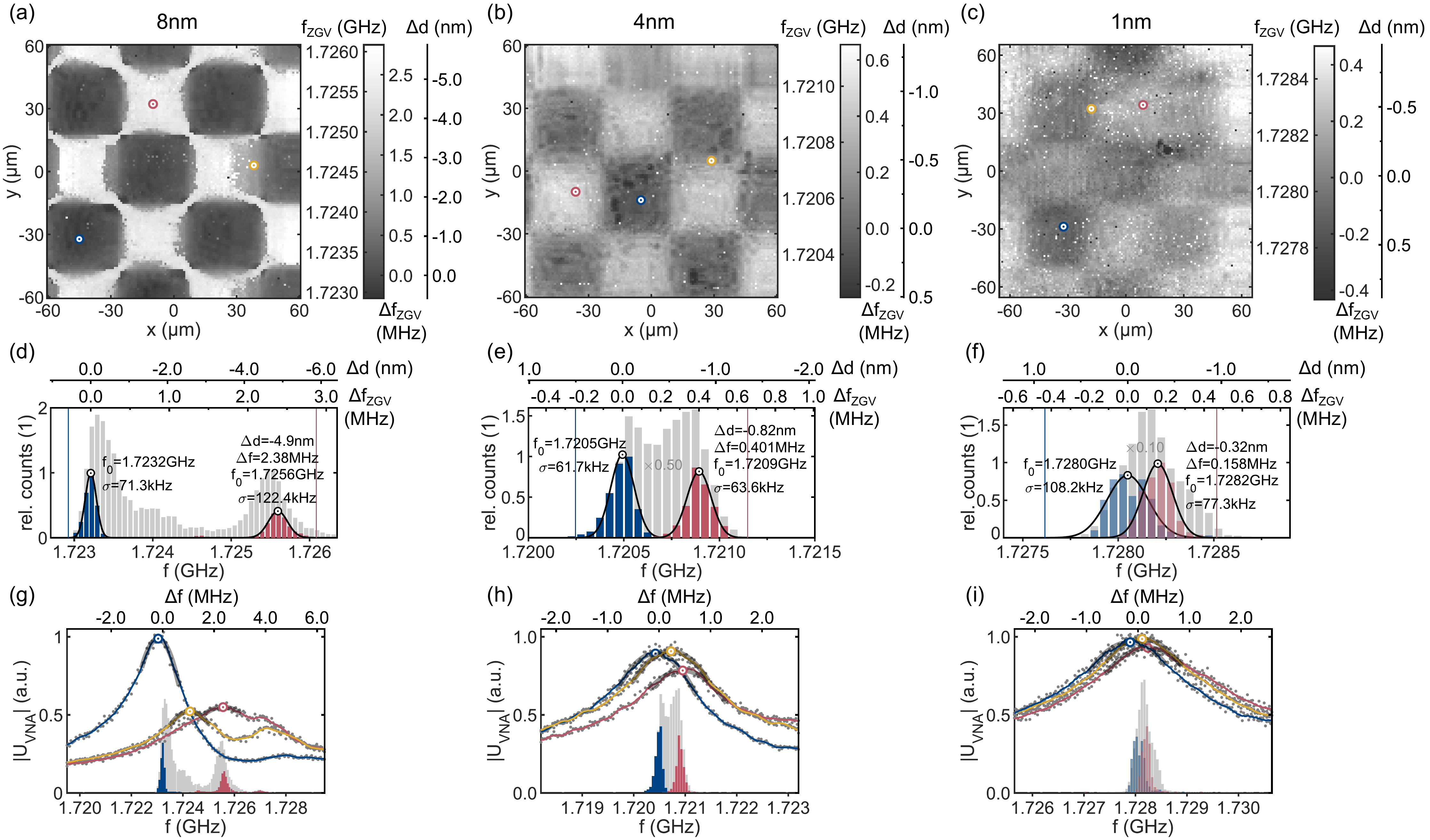}
	\caption{\textbf{ZGV-resonance based nanometric height-imaging I.} (a-c) Images of checkered resonators with different nominal step heights generated from measured ZGV-resonance center-frequencies. Lower regions appear brighter in the image. The additional $\Delta{}d$-axis on the right hand sides of the colorbars were calculated based on \equref{eq:freqThick} (d-f) Histograms of extracted ZGV-center frequencies extracted from five lowered squares (red bars; bright squares in (a-c)) and five squares with unchanged SiN-layer thickness (blue bars; dark squares in (a-c)) and according Gaussian-fits (black solid lines). The ZGV-frequency distribution of the entire image is indicated by the grey bars in the background (they are scaled with the specified factors to fit to the graph). (g-i) Example spectra from lowered regions in (red color), from regions with original SiN-height (blue color) and from transition regions (yellow color). The locations are marked in (a-c) with colored dots/circles.\label{fig:checkered}}%
\end{figure*}
To demonstrate sub-\SI{10}{nm} quantitative height mapping we 2D-scanned the checkered resonators on a regularly spaced grid with a \SI{1}{\micro\metre} lateral increment. The extracted ZGV-resonance frequencies are plotted as greyscale images in \figref[(a-c)]{fig:checkered}. The checkerboard pattern is clearly visible for the \SI{8}{nm}-, and the \SI{4}{nm}-case, and can still be recognized for the lowest nominal step height of \SI{1}{nm}. This is quite remarkable, as the lattice constants of hexagonal SiN are of comparable size ($a\!=\!\SI{0.76}{nm}$, $b\!=\!\SI{0.29}{nm}$, in \href{https://doi.org/10.17188/1316765}{The Materials Project}). As for the stepped case shown before, \equref{eq:freqThick} was used to transform ZGV-frequency shifts into height-change (see $\Delta{}d$-axis in the graphs). We emphasize that the range of the colorbars in \figref[(a-c)]{fig:checkered} was systematically chosen based on the statistics of the measured data (see Methods section for details) to reveal as much detail as possible in the images. On the other hand, by tweaking the colobar limits it is easily possible to obtain images which perfectly reproduce checkerboard patterns. However, by that, the details at the edges in the \SI{8}{nm}-case (\figref[(a)]{fig:checkered}) are lost for instance.

To estimate the step height from the 2D-scans we generated the histograms depicted in \figref[(d-f)]{fig:checkered} by collecting ZGV-frequencies from the centers of the ablated (red bars) and non-ablated (blue) squares (more details can be found in the Methods section). From the difference in the ZGV-frequency expectation values (see Gaussian-normal distribution fits as solid black lines) the step heights can be estimated. The retrieved values are systematically lower as those from the stepped resonator scans (see \tabref{tab:height_all}). While the results for the \SI{8}{nm}-case are comparable, they strongly deviate for the \SI{4}{nm}-, and \SI{1}{nm}-case. This may be partially explained by the larger point-spread functions ($\Delta{}y$) in these cases.
\begin{table}[h!btp]
	\renewcommand{\arraystretch}{1.15}
	\caption{Extracted step heights (in units of $(\si{nm})$) from stepped and checkered resonators and ratio between stepped and checkered step heights.} \label{tab:height_all}
	\centering
	\begin{tabular}{  c c c c }
		\toprule
		nominal  &  stepped & checkered & ratio\\	
		\midrule
		8 & 6.2 & 4.9 & 1.3\\			
		4 & 1.9 & 0.8 & 2.4\\		
		1 & 1.2 & 0.3 & 4.0\\
		\bottomrule
	\end{tabular}
\end{table}

By removing outliers from the datasets (mainly the white spots in \figref[(a-c)]{fig:checkered}), 3D topographic images as shown in \figref[(a-c)]{fig:checkered3D} can be generated. They provide a more detailed view of the sample surfaces, and show, that also for the checkered samples, the steps in the \SI{4}{nm}- and \SI{1}{nm}-samples are much smoother than in the \SI{8}{nm}-case.

\begin{figure*}[h!tbp]
	\centering
	\includegraphics[width=1.0\textwidth]{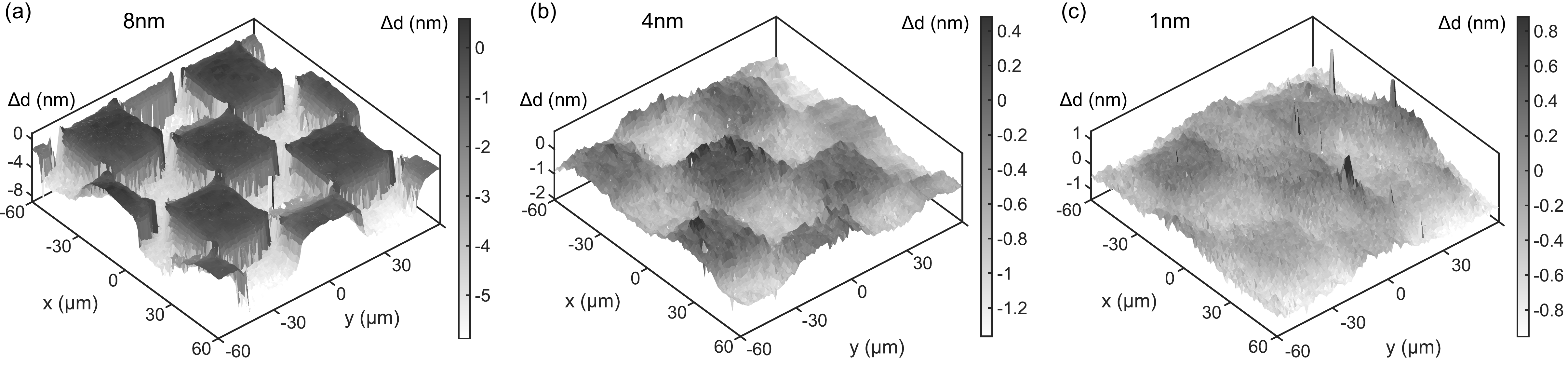}
	\caption{\textbf{ZGV-resonance based nanometric height-imaging II.} (a-c) 3D-Images of checkered resonators with different nominal step heights generated from measured ZGV-resonance center-frequencies. Outliers have been removed from the datasets.\label{fig:checkered3D}}%
\end{figure*}

\subsection*{Comparison to AFM measurements}
\begin{figure}[h!tbp]
	\centering
	\includegraphics[width=0.5\textwidth]{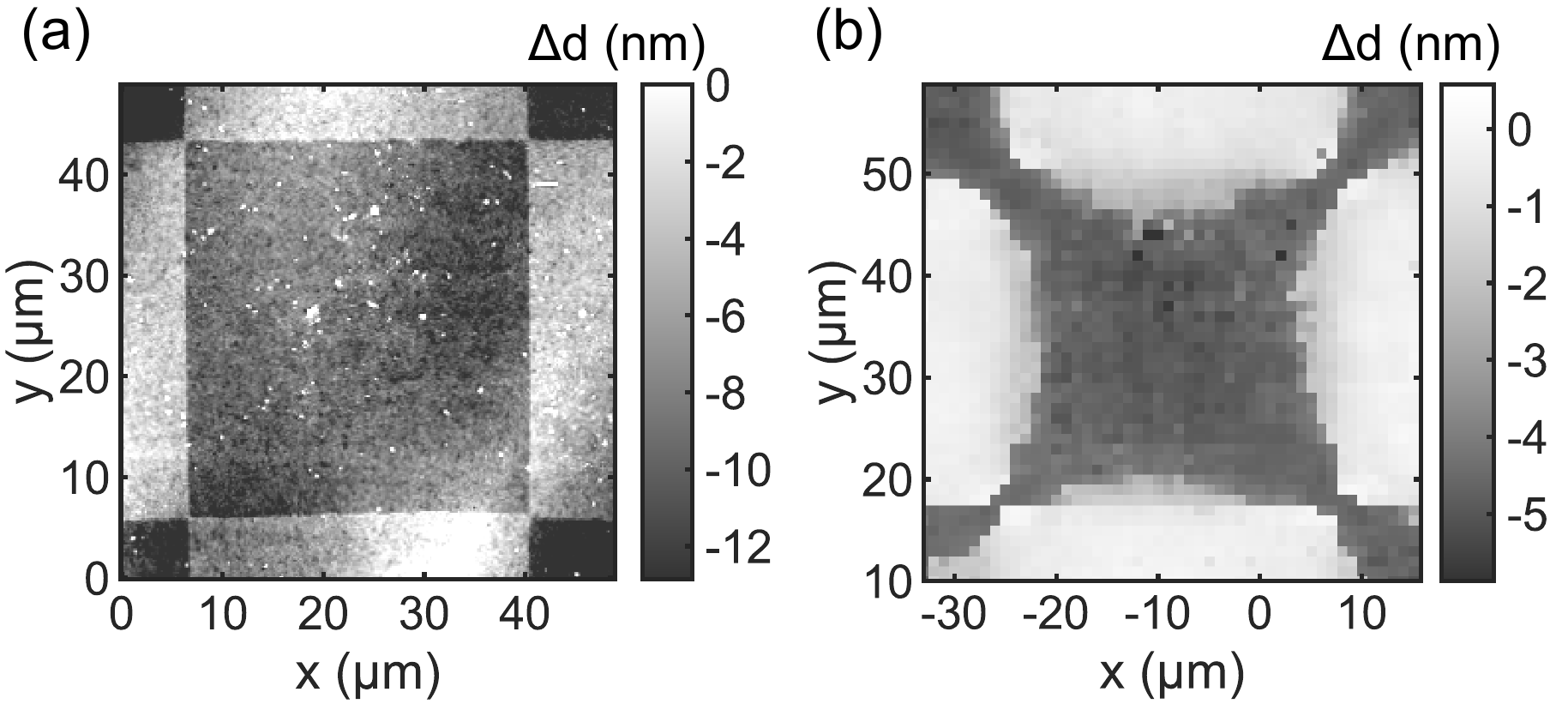}
	\caption{\textbf{AFM-imaging vs.\ ZGV-resonance based LUS imaging} (a) AFM-image of a \SI{8}{nm} stepped resonator. (b) ZGV-resonance based LUS-image of a similar region with the same lateral dimensions in a \SI{8}{nm} stepped resonator (Taken from the center region of \figref[(a)]{fig:checkered}. Note that the colorbar is inverted compared to the previous image)\label{fig:AFM_LUS}}%
\end{figure}
\begin{table}[h!btp]
	\renewcommand{\arraystretch}{1.15}
	\caption{Extracted step heights in resonators with a nominal top-layer step height of \SI{8}{nm}.} \label{tab:height_8nm}
	\centering
	\begin{tabular}{  c c c }
		\toprule
		method &  pattern & step height $(\si{nm})$ \\	
		\midrule
		AFM & checkered     & 6.4 \\			
		FreDomLUS & stepped 		& 6.2 \\		
		FreDomLUS & checkered 	& 4.9 \\
		\bottomrule
	\end{tabular}
\end{table}
To verify the ZGV-resonance based height imaging results, we performed atomic force microscopy (AFM) reference measurements in a \SI{8}{nm} checkered resonator structure. \Figref{fig:AFM_LUS} shows a comparison between the AFM-image and FreDomLUS-results, taken from a central region of \figref[(a)]{fig:checkered} with the same lateral dimensions as the AFM image (Note that the colorbar is reversed in this image). As expected, the lateral resolution of AFM outperforms the FreDomLUS-method, and the sharp edges of the checkerboard-squares are reproduced excellently. Contrary, the relatively large pointspread function of the ZGV-resonance based method  (around $\SI{3.2}{\micro\metre}$; see $\Delta{}y$ in \figref[(a)]{fig:steppedHeight}) leads to a blurring of the edges, especially in the corner regions of the squares. We also determined the step height from the AFM measurement from line profiles across the vertical edges. Their average gives a step height of \SI{6.4}{nm} which agrees excellently with the \SI{6.2}{nm} FreDomLUS result obtained from the stepped resonator scan (see \tabref{tab:height_8nm}). This confirms the observed deviation from the nominal step-height of \SI{8}{nm} by the FreDomLUS-method, and suggests that also resonators with \SI{4}{nm} and \SI{1}{nm} nominal top-layer height variation are actually thinner, and the measured LUS step-heights of \SI{1.6}{nm} and \SI{1.2}{nm} are reliable (see \figref{fig:steppedHeight} (b) and (c)). 

\section*{Discussion}
In this study, we investigated the potential of a non-contact laser-based acoustic microscopy technique for quantitative nanometric height imaging. It synergizes the advantages of the FreDomLUS-measurement technique and the inverse thickness scaling of elastic plate resonances. While the FreDomLUS-technique allows us to non-destructively probe local acoustic response spectra up to GHz frequencies with a virtually arbitrary fine frequency resolution, the employed local zero-group velocity ZGV-resonances lead to sharp, high-quality factor peaks that inversely scale with plate thickness. For demonstration experiments, we patterned the top layer of \SI{1.7}{GHz} solidly mounted BAW-resonators by introducing sub \SI{10}{nm} height-changes in its top layer. The main contributions of this work are: (1) We have shown that ZGV-resonances exist in multi-layered BAW-resonator structures and that they can be probed by the FreDomLUS-technique at GHz frequencies. (2) We compared theoretical dispersion curves obtained from a spectral collocation method that incorporates electro-elastic coupling and found excellent agreement with measured response spectra. (3) We have shown, how these dispersion curves can be used to calibrate a linear model that allows us to calculate thickness changes from a change in ZGV-frequency. (4) We estimated the lateral resolution of ZGV-based imaging technique by measuring their step-edge function and obtained values in the \SI{}{\micro\metre}-range. (5) We have demonstrated quantitative sub \SI{10}{nm} height-imaging.

Please note that the obtained method does not rely on the used frequency domain LUS-technique, and can, in principle, also be performed by time-domain laser ultrasonic approaches. However, we point out two FreDomLUS-specific advantages, which makes the technique more suitable for that specific application: First, the time-harmonic excitation with an intensity modulated laser source in combination with narrow-band phase-sensitive detection enables the usage of low excitation powers in the mW-range, instead of the typical MW peak powers in pulsed laser sources typically used in time-domain LUS. This allows us to investigate rather thin micron-scaled specimens. Second, the frequency domain-approach is inherently matched to the measurement problem, that is, to measure narrow band acoustic resonances in the frequency domain which inversely scale with thickness. Especially when small relative frequency changes which require fine frequency resolution are involved (For example, in our \SI{1}{nm}-case, a relative frequency change of \SI{0.03}{\percent} was measured). In time-domain measurements, this simultaneously requires fine temporal sampling and long recording times of the exponentially decaying resonance signals, which sets high demands on signal to noise ratio and thus on the used instrumentation.

Clearly, the lateral resolution of the ZGV-based acoustic imaging cannot compete with AFM (\unit{\micro\metre} vs.\ sub-\unit{nm}), and is ultimately limited by the wavelength of the employed plate-resonance. This is set by the elastic properties and the layer-structure of the studied specimen. However, unlike in AFM, the field of view of the method is larger and may extend over several centimeters. Although we did not study the depth-sensitivity of the method systematically, it is instructive to consider the example spectra taken from the checkered sample with a nominal height of \SI{1}{nm} in \figref[(i)]{fig:checkered}: While in that case the height-variations can still be detected (\figref[(c)]{fig:steppedHeight} and \figref[(c)]{fig:checkered}), the single spectra suggest, that a combination of SNR and the quality factor of the used resonance (i.e. the peak width) limits the height detection sensitivity. This is strongly dependent on the investigated material system.

The investigations have been carried out on devices which have been designed to efficiently sustain a resonant mode with a frequency close to the investigated ZGV-resonance (see \figref[(d)]{fig:Figure_ArticleVersion_1}) and are thus inherently well-suited for the presented study. Thus, we expect that the method is applicable for layer thickness characterization of other micro-electromechanical systems (MEMS), in particular (but not only) different electroacoustic resonator designs like free-standing resonators (FBARs) for instance. In this regard, we note that the demonstrated local lateral sensitivity of the method may be used to study unwanted lateral variations of the electroelastic properties that might deteriorate device performance. While an artificial thickness change was chosen here for demonstration purposes, also varying elastic properties due to microstructural variations or any kind of defects fall into this category. At the same time we expect that the method can be applied for imaging of defects, film-thickness variations, or debonding in microelectronic devices, similar to scanning acoustic microscopes (SAMs).

While here nanometric thickness variations of a \SI{205}{nm} thick top-layer was investigated, the wave motion of the utilized ZGV-resonance extends several micrometers into deeper layers of the stack (see ZGV-mode-shapes drawn as red solid lines in \figref[(a)]{fig:Figure_ArticleVersion_1}). In future work, we want study the sensitivity of the method with respect to thickness variations of sub-surface layers and test it on non-resonant structures as well.

\section*{Methods}\label{sec:methods}
\subsection*{Sample fabrication and properties} 
Solidly mounted BAW resonators were fabricated with state-of-the-art micro-fabrication techniques on a \SI{725}{\micro\metre} thick silicon wafer (see \figref[(a)]{fig:Figure_ArticleVersion_1}). They consist of an acoustic Bragg-mirror composed of five alternating layers of SiO$_2$ and W, followed by a ScAlN9 piezoelectric layer between two sub-micron electrodes. The stacks are quadratic with a lateral length of \SI{150}{\micro\metre} and are passivated with a \SI{205}{\nano\metre} thick SiN layer on top. The exact stack composition is specified in \tabref{tab:stack}. The top layers were patterned by standard optical lithography and ion beam etching to produce artificial height-change patterns, either in the form of a single step ('stepped resonator') or as a checkerboard pattern ('checkered resonator') with a period of \SI{39}{\micro\metre} (\figref[(b)]{fig:Figure_ArticleVersion_1}). The nominal depths of the ablated regions amount to \SI{8}{nm}, \SI{4}{nm}, or \SI{1}{nm}, each on a separate wafer.
For forward calculations of guided-wave dispersion curves with the SCM-method\cite{kiefer_electroelastic_2025} all materials in the BAW-stack are considered elastically isotropic, except for the piezoelectric AlN- and ScAlN9-layers, which have hexagonal crystal symmetry with their $c$-axis oriented along the $z$-direction. The used elastic material constants and material densities are specified in \tabref{tab:mat_iso} and \tabref{tab:mat_hex}. The piezoelectric- and dielectric constants of AlN and ScAlN9 are specified in \tabref{tab:mat_hex_piezo}. Perfect conductivity was assumed for the metallic electrodes in the SCM model.

\begin{table}[h!btp]
	\renewcommand{\arraystretch}{1.15}
	\caption{Layer-composition of BAW-resonator samples.} \label{tab:stack}
	\centering
	\begin{tabular}{  c c r }
		\toprule
		layer \# & material & nominal thickness $(\si{nm})$ \\	
		\midrule
		1 & Si 				& 725000 \\	
		2 & SiO$_2$ 		& 670 \\
		3 & AlN 			& 20 \\		
		4 & W 				& 670 \\
		5 & SiO$_2$ 		& 685 \\
		6 & AlN 			& 20 \\								
		7 & W 				& 870 \\
		8 & SiO$_2$ 		& 440 \\
		9 & AlN 			& 20 \\
		10 & AlCu			& 200 \\
		11 & W 				& 325 \\
		12 & ScAlN9			& 1210 \\
		13 & W 				& 42 \\
		14 & TiW 			& 20 \\
		15 & AlCu			& 165 \\
		16 & SiN 			& 205 \\
		\bottomrule
	\end{tabular}
\end{table} 
\begin{table}[h!btp]
	\renewcommand{\arraystretch}{1.15}
	\caption{Stiffness tensor components ($C_{ij}$) and density ($\rho$) of the isotropic materials used in plate-dispersion calculations.} \label{tab:mat_iso}
	\centering
	\begin{tabular}{  c c c c }
		\toprule
		material & $C_{11} (\si{GPa})$ &  $C_{44} (\si{GPa})$ & $\rho (\si{kg.m^{-3}})$\\	
		\midrule
		Si 				& 165.70 & ~79.56 & ~2300 \\	
		SiO$_2$ 		& ~74.32 & ~33.62 & ~2207 \\	
		W 				& 522.40 & 141.73 & 19758 \\
		AlCu			& 111.28 & ~26.11 & ~2700 \\
		TiW				& 383.97 & 119.43 & 14200 \\
		SiN				& 164.86 & ~58.01 & ~2471 \\
		\bottomrule
	\end{tabular}
\end{table} 
\begin{table}[h!btp]
	\renewcommand{\arraystretch}{1.15}
	\caption{Stiffness tensor components ($C_{ij}$) and density ($\rho$) of the hexagonal materials used in plate-dispersion calculations.} \label{tab:mat_hex}
	\centering
	\begin{tabular}{  c c c c c c}
		\toprule
		material & $C_{11} (\si{GPa})$ &  $C_{12} (\si{GPa})$ &  $C_{13} (\si{GPa})$ &  $C_{44} (\si{GPa})$ & $\rho (\si{kg.m^{-3}})$\\	
		\midrule
		AlN 	& 396.00 & 137.00 & 108.00 & 116.00 & 3255 \\	
		ScAlN9 	& 366.50 & 141.84 & 115.29 & 108.50 & 3280 \\
		\bottomrule
	\end{tabular}
\end{table} 

\begin{table}[h!btp]
	\renewcommand{\arraystretch}{1.15}
	\caption{Piezoelectric coupling tensor components ($e_{ij}$) and relative permittivity matrix components ($\epsilon^\mathrm{r}_{ij}$) of thehexagonal materials used in plate-dispersion calculations.} \label{tab:mat_hex_piezo}
	\centering
	\begin{tabular}{  c c c c c c}
		\toprule
		material & $e_{15} (\si{C.m^{-2}})$ &  $e_{31} (\si{C.m^{-2}})$ &  $e_{11} (\si{C.m^{-2}})$ & $\epsilon^\mathrm{r}_{33} (1)$ &  $\epsilon^\mathrm{r}_{33} (1)$\\	
		\midrule
		AlN 	& -0.313 & -0.593 & 1.471 & 10.72 & 10.31\\	
		ScAlN9  & -0.300 & -0.614 & 1.617 & 10.72 & 12.36 \\
		\bottomrule
	\end{tabular}
\end{table}

\subsection*{Frequency domain laser ultrasound measurements}
The $\nicefrac{1}{e}$-excitation laser spot size amounted to approximately \SI{2.4}{\micro\metre} in all measurements and its peak power was set to \SI{250}{mW}. The complex valued acoustic response spectra $\tilde{U}_\mathrm{VNA}^\mathrm{sig}$ were recorded with a frequency step size of either \SI{25}{kHz} or \SI{50}{kHz} and a VNA-bandwidth of \SI{100}{Hz}. In addition, at each point, a background spectrum $\tilde{U}_\mathrm{VNA}^\mathrm{ref}$ was recorded by blocking the excitation laser with a mechanical shutter, but otherwise using the same settings. For further processing, the difference $\tilde{U}_\mathrm{VNA}\!=\!\tilde{U}_\mathrm{VNA}^\mathrm{sig}-\tilde{U}_\mathrm{VNA}^\mathrm{ref}$ was used. All spatial scans were performed with a \SI{1}{\micro\metre} lateral step size.

\subsection*{Data processing details}
ZGV-center frequency extraction: For that we use the absolute value of the background-subtracted signals $|U_\mathrm{VNA}(f)|$ (see for example \figref[(b)]{fig:Stepped_8nm}). The curves are smoothed with a 2$^\mathrm{nd}$-order polynomial Savitzky-Golay filter using a window size of 17 data points. These resulting traces were fitted with 2$^\mathrm{nd}$-order polynomials of the form $|U_\mathrm{VNA}(f)|\!=\!|U_\mathrm{VNA}|_\mathrm{max}-k(f-f_\mathrm{ZGV})^2$ in a \SI{1.5}{MHz} broad frequency band centered at the trace maximum. This yields the ZGV center frequency $f_\mathrm{ZGV}$ which was used for further processing (imaging).

Statistical analysis of 2D-maps in \figref[(a-c)]{fig:checkered} and setting of colorbar limits: To create the histograms in \figref[(d-e)]{fig:checkered} we manually selected quadratic regions of interest (ROIs) in the center of the checkerboard squares shown \figref[(a-c)]{fig:checkered}. The lateral lengths of the ROIs (\SI{18}{\micro\metre} for the \SI{8}{nm} and \SI{4}{nm}-cases, and \SI{12}{\micro\metre} for the \SI{1}{nm}-case) were set smaller than the checkerboard-square length to avoid influence of the steps, and get statistical representation of the squares with original thickness (black-squares in \figref[(a-c)]{fig:checkered} and blue bars in \figref[(d-f)]{fig:checkered}) and the ablated regions (white-squares in \figref[(a-c)]{fig:checkered} and red bars in \figref[(d-f)]{fig:checkered}). The obtained histograms were fit with a Gaussian normal distribution (black solid lines in \figref[(d-f)]{fig:checkered}). It's expectation value was used as an estimate for the average ZGV-frequency in the non-ablated and ablated regions ($f_\mathrm{ZGV,0}^\mathrm{lo}$, $f_\mathrm{ZGV,0}^\mathrm{hi}$). Their difference was used to estimate the step-heights $\Delta{}d$ from the images via \equref{eq:freqThick}. The colorbar ranges in \figref[(a-c)]{fig:checkered} were set to cover intervals of $(f_\mathrm{ZGV,0}^\mathrm{lo}-4\sigma^\mathrm{lo}, f_\mathrm{ZGV,0}^\mathrm{hi}+4\sigma^\mathrm{hi})$ for each case, where $\sigma^\mathrm{lo}$ and $\sigma^\mathrm{hi}$ are the standard deviations of the Gaussian fits. These ranges are indicated by the vertical blue and red lines in \figref[(d-f)]{fig:checkered}.

For the 3D-topography images in \figref[]{fig:checkered3D} we automatically removed outliers from the datasets by setting a threshold for the (polynomial) fit root mean square error (RMSE) value ($\mathrm{RMSE} < \mu_\mathrm{RMSE}+ 6\sigma_\mathrm{RMSE}$) and the ZGV-peak magnitude ($|{U}_\mathrm{VNA}^\mathrm{max}| < \mu_{|{U}_\mathrm{VNA}^\mathrm{max}|}+ 5\sigma_{|{U}_\mathrm{VNA}^\mathrm{max}|}$).

\subsection*{Atomic force microscopy}
The AFM-images were recorded with a Keysight 5500 AFM in tapping mode using NanoWorld NCHR standard tapping mode AFM probes and a pixel size of \SI{200}{nm}. The raw data was processed with the scanning probe microscopy image processing software \emph{Gwyddion} (\url{https://gwyddion.net/}): Background correction was done by mean plane subtraction followed by 2$^\mathrm{nd}$-order polynomial correction. Line artifacts were corrected by the software's automatic line artifact correction routine ('align rows').

The average step height (\tabref{tab:height_8nm}) was extracted from the two vertical steps shown in \figref[(a)]{fig:AFM_LUS} as they are approximately orthogonal to the tip movement direction ($x$-direction). For this purpose horizontal line profiles across the two steps with a length of \SI{5}{\micro\metre} were extracted (excluding a region within a distance of less than \SI{2}{\micro\metre} to the top and bottom horizontal steps) and the average $z$-value was calculated for the upper and lower regions, starting with positions at least \SI{0.55}{\micro\metre} apart from the step. From these, a set of step-heigths was calculated and the obtained distribution was fit with a Gaussian normal distribution that yielded an average step-height of \SI{6.42}{nm} with a standard deviation of \SI{1.19}{nm}.

\section*{Data Availability}
The data that support our findings in this paper are available from the corresponding author upon reasonable request.

\begin{acknowledgments}
This work was funded by the Austrian research funding association (FFG) under the scope of the COMET program within the research project "Photonic Sensing for Smarter Processes (PSSP)" (Contract No.\ 871974) and by the project “HyperMAT” by the federal government of Upper Austria and the European Regional Development Fund (EFRE) in the framework of the EU-program IBW/EFRE \& JTF 2021-2027.
\end{acknowledgments}

\section*{Author Contributions}
	\textbf{Martin Ryzy:} %
		conceptualization (equal);
		methodology (lead);
		software (lead);
		validation (equal);
		formal analysis (lead);
		investigation (equal);
		data curation (lead);
		writing – original draft (lead); 
		writing – review and editing (equal).		
		project administration (lead);
		visualization (lead);
		funding acquisition (equal).
	\textbf{Guqi Yan:} 
		methodology (equal);
		formal analysis (equal);
		investigation (lead);
		data curation (supporting);
		writing – original draft (equal);
		writing – review and editing (equal);
		visualization (equal).
	\textbf{Istv\'{a}n Veres:} 
		methodology (equal);
		resources (equal);
		writing – review and editing (supporting);
		supervision (lead).
	\textbf{Thomas Berer:}
		methodology (supporting);
		resources (equal).
		writing – review and editing (supporting).
		supervision (equal).
	\textbf{Ivan Ali\'{c}:}
		investigation (supporting);
		writing – review and editing (supporting).
	\textbf{Clemens Grünsteidl:}
		methodology (equal);
		software (supporting);
		writing – review and editing (supporting);
		supervision (supporting).
	\textbf{Georg Watzl:}
		validation (supporting);
		writing – review and editing (supporting).
	\textbf{Georg Gramse:}
		resources (supporting);
		writing – review and editing (supporting).
	\textbf{Susanne Kreuzer:}
		conceptualization (equal);
		resources (lead);
		writing – review and editing (equal);
		supervision (supporting).
		funding acquisition (equal).

\section*{Competing interests}
The authors declare that they have no competing financial interests or personal relationships that could have appeared to influence the work reported in this paper.

\bibliographystyle{apsrev4-2}
\bibliography{MyCollection_NanoThickImage}

\end{document}